\documentclass[conference]{IEEEtran}
\usepackage{graphicx}

\usepackage{amsmath}

\ifCLASSINFOpdf

\else
  
\fi

\begin{document}

\title{From EEG Cleaning to Decoding: \\ The Role of Artifact Rejection in MI-based BCIs}

\author{
\IEEEauthorblockN{
Davoud Hajhassani\textsuperscript{1},
Bruno Aristimunha\textsuperscript{1},
Paul-Adrien Graignic\textsuperscript{1},
Apolline Mellot\textsuperscript{1},
Lionel Kusch\textsuperscript{1}, \\
Arnaud Delorme\textsuperscript{2},
Thomas Semah\textsuperscript{1},
Arnault H. Caillet\textsuperscript{1,*}
}
\IEEEauthorblockA{
\textsuperscript{1}\textit{Yneuro}, Paris, France\\
\textsuperscript{2}\textit{SCCN, INC, SDSC, University of California San Diego}, La Jolla, USA\\
\textsuperscript{*}\texttt{arnault@yneuro.com}
}
}

\maketitle

\begin{abstract}
Motor imagery (MI) BCIs are sensitive to EEG artifacts, yet the practical impact of automated artifact rejection on downstream MI decoding performance remains unclear. 
While most work focuses on decoder design, the contribution of data curation, particularly automated rejection policies, has received comparatively less attention, despite its importance for robust machine learning pipelines. 
Here, we propose Fast Automatic Artifact Rejection (FAAR), a lightweight method that computes a compact set of artifact-sensitive features, derives an epoch-level Signal Quality Index, adaptively selects rejection thresholds, and automatically rejects contaminated epochs without requiring prior knowledge of artifact types or manual threshold tuning.
We evaluate FAAR on 13 publicly available MI datasets and compare it to a no-rejection baseline, AutoReject, and Isolation Forest. We show rejection effects are strongly subject- and regime-dependent, with the largest gains in low-baseline/low-SNR conditions, so it should be used adaptively. FAAR reduces inter-subject performance variability, an important property for MI-BCI reliability and BCI-illiteracy, without aggressive data removal. Finally, FAAR’s lightweight and fully automated thresholding yields consistent rejection behavior across offline curation, training, and online filtering, and supports real-time BCI constraints.

\end{abstract}


\IEEEpeerreviewmaketitle

\section{Introduction}

Motor imagery (MI) is a central paradigm in brain–computer interface (BCI) research because it enables users to control external devices solely by imagining movement, without engaging peripheral nerves \cite{wolpaw1991cursor}. 
This has enabled health applications ranging from motor rehabilitation to assistive control \cite{khan2020mibci_review,pan2026fiveclass}, and also motivates augmented and virtual reality BCI use cases where decoding user intent in-the-loop can support frictionless interactions, including authentication-oriented scenarios in augmented environments \cite{Caillet2025NeuroBioAuthPatent}. 
MI-based BCIs are commonly built from electroencephalography (EEG), which offers high temporal resolution with relatively low-cost, non-invasive and portable acquisition \cite{abiri2019bci_paradigms}.
Yet, EEG recordings are often contaminated by non-cerebral electrical activity (\emph{i.e.}, artifacts) originating from physiological sources (\emph{e.g.}, eye movements, muscle and cardiac activity) and from instrumental or environmental factors (\emph{e.g.}, impedance mismatches and electromagnetic interference) \cite{bashashati2007survey}.

Artifacts can substantially degrade signal-to-noise ratio (SNR), which is particularly detrimental in MI where discriminative rhythms can be subtle. 
In practice, artifacts can mask task-relevant modulations and induce false-positive predictions, reducing the reliability of MI-based BCIs. 
Moreover, because artifacts are nonlinear and nonstationary, handling them without distorting neuronal activity remains challenging \cite{mumtaz2021artifact_review}.
This motivates integrating, into EEG analysis pipelines, artifact-handling stages that are effective while remaining operationally compatible with real-time BCI constraints. Artifact handling is commonly framed as \emph{artifact correction} (suppress artifacts while preserving EEG) versus \emph{artifact rejection} (exclude contaminated epochs from subsequent analysis) \cite{hajhassani2026riemannian_potato}. 
A wide range of correction methods has been proposed \cite{mumtaz2021artifact_review,islam2016artifact_review, jiang2019artifact_review_sensors}, and several studies have investigated how artifact correction influences downstream performance in MI-based BCIs, including trade-offs between denoising and distortion of task-relevant neural rhythms \cite{frolich2015artifact_types_mi, adolf2024processing_techniques, geng2022improved_feature_extraction, mohammadi2021eog_mi,tobonhenao2022subject_dependent_artifact}. 
By contrast with correction, the impact of \emph{artifact rejection} on MI-based BCI performance has been less systematically characterized. 

In practice, artifact rejection methods have shifted from manual expert review \cite{jung2000bss_artifacts} (time-consuming, subjective, and incompatible with scale and real time \cite{delorme2023eeg_left_alone}) to automated pipelines (\emph{e.g.}, EEGLAB, EEGPrep, MNE) \cite{delorme2004eeglab,eegprep,gramfort2013mnepython}, that reduce manual effort and improve scalability and reproducibility.
Most semi-automated implementations rely on fixed epoch-rejecting thresholds on signal-quality metrics (\emph{e.g.}, peak-to-peak amplitude), as in FASTER \cite{nolan2010faster}. 
However, EEG and artifact statistics vary widely across subjects, sessions, and datasets, so fixed thresholds rarely transfer and usually require dataset-specific recalibration, limiting scalable curation and real-time use. 
Autoreject (AR) automates threshold selection via cross-validation and Bayesian optimization \cite{jas2017autoreject}, but at higher computational cost that limits large-scale processing or online use. 
Isolation Forest (IF) provides a faster automated alternative \cite{zhang2024isolationforest_eeg}, but can be aggressive in epoch removal. We include AR and IF as reference methods in our comparisons.

A complementary family of approaches uses Riemannian geometry: epochs are mapped to covariance matrices and scored by a Signal Quality Index (SQI) relative to a clean reference set \cite{barachant2013riemannian_potato,barthelemy2019potato_field}. Reference-based SQIs with adaptive thresholding can outperform AR and IF in P300 settings \cite{hajhassani2026riemannian_potato,hajhassani2024riemannian_p300}, but transferring across contexts remains difficult because the definition of the reference set and the quality representation often depend  on montage, artifact structure, and task-specific choices.

In this paper, we propose \emph{Fast Automatic Artifact Rejection (FAAR)}, a lightweight, real-time compatible artifact rejection method designed to generalize across EEG recordings and be compatible with online settings. We evaluate FAAR on 13 public MI datasets and compare it to AutoReject (AR), Isolation Forest (IF), and a no-rejection baseline, reporting decoding performance, rejection rate, and robustness across subjects.

\section{Methods and Materials}
\subsection{Fast Automatic Artifact Rejection}
FAAR is a lightweight artifact rejection method that computes an epoch-level SQI from (i) a compact set of artifact-sensitive features and (ii) a subject-specific clean reference estimated from the recording itself, then (iii) selects an adaptive SQI threshold from the SQI distribution to automatically label epochs as retained or rejected. 

\subsubsection{Signal Quality Features}
FAAR computes five complementary descriptors that contribute to the SQI computation and jointly capture major artifact signatures by being broadly sensitive to common EEG artifact families, complementary across time/frequency/statistics, and analytic and low-cost to compute for real-time use \cite{fickling2019eegqi}. For each channel and short window we compute:

\begin{itemize}
    \item Band-limited spectral magnitude (8--32 Hz). Targets abnormal power elevation in MI-relevant $\alpha/\beta$ bands; computed as mean $|$FFT$|$ coefficients in-band.
    \item RMS amplitude. Targets global amplitude/variance outliers (\emph{e.g.}, pops, saturation, ocular/motion); computed as window RMS \cite{kappenman2010impedance}.
    \item Maximum temporal gradient. Targets sharp transients/step changes missed by band-power (\emph{i.e.}, high-amplitude/-frequency artifacts, motion transients, electrode shifts); computed as max $|\Delta x|$ between consecutive samples.
    \item Zero-crossing rate. Coarse frequency proxy for low/high-frequency content (\emph{e.g.}, EMG$\uparrow$, drift$\downarrow$); computed as sign-change rate \cite{sun2020complexity_review}.
    \item Kurtosis. Targets non-Gaussian impulsive/bursty outliers not well
captured by variance alone; computed as normalized 4th central moment \cite{najim2004stochastic,delorme2007artifact_ica}.
\end{itemize}

\subsubsection{Reference Data Extraction (self-calibration)}
To avoid manually defining dataset-specific thresholds, FAAR derives a subject/session-specific \emph{clean reference} directly from the recording itself. To do so, FAAR quantifies channel-wise signal variance by computing RMS values across short windows (\emph{e.g.}, 1-s) and standardizing them per channel (z-scoring across windows). A subset of \emph{clean} windows is then automatically selected as reference  by modeling the standardized RMS distribution with a truncated-Gaussian inlier criterion \cite{chang2020asr_eval}. To remain usable in noisy recordings, reference selection retains windows with only a small fraction of outlier (\emph{bad}) channels, ensuring sufficient reference data even under degraded acquisition conditions. 

The resulting reference distribution is used to normalize subsequent feature values; this self-calibration method avoids manual recalibration of SQI normalization and relying on fixed simulated \emph{reference} data, which can be suboptimal given inter-subject variability \cite{fickling2019eegqi}.
In online deployment, this reference can be initialized from an initial warm-up buffer (\emph{e.g.}, the first seconds), or from a previously learned subject/device reference (\emph{e.g.}, enrollment or prior sessions); it can then be maintained as a rolling (or exponentially-forgetting) estimate and updated incrementally in real time as new clean windows are observed.

\begin{figure}
    \centering
    \includegraphics[width=1\linewidth]{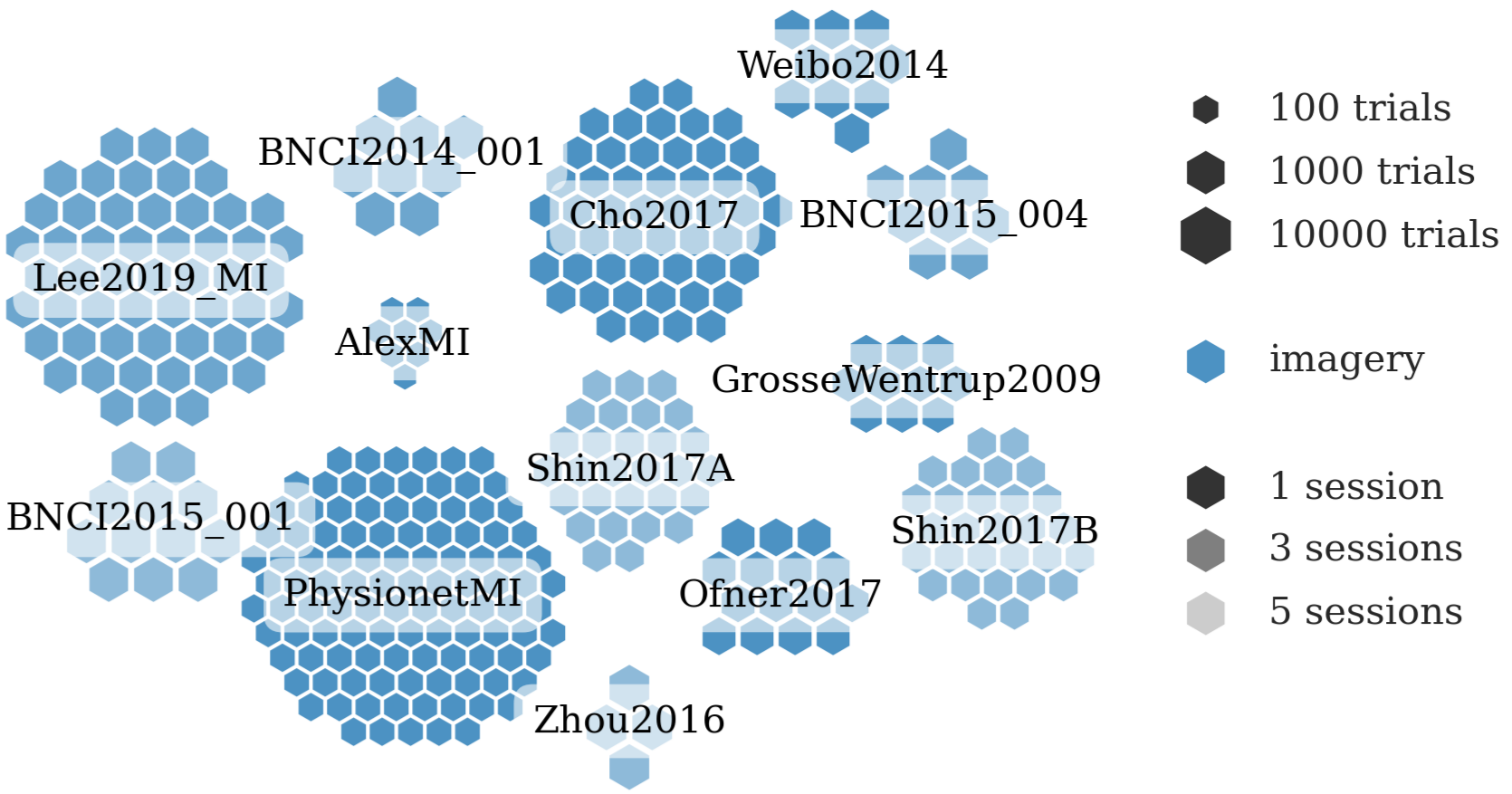}
    \caption{A visual overview of the MI datasets included in this study. Hexagon clusters indicate the relative number of trials per dataset, while shading encodes the number of recording sessions. All datasets are publicly available through the MOABB framework \cite{aristimunha2025moabb}.}
    \label{datasets}
\end{figure}

\subsubsection{Scoring and adaptive thresholding}
Rather than applying fixed hand-set thresholds to each signal quality feature, FAAR adaptively scores feature deviations relative to the subject/session clean reference (c.f. previous section). Concretely, for each feature and channel, values are standardized with respect to the reference distribution (z-scoring) and mapped to an ordinal severity score (\emph{e.g.}, four levels from “within reference” indicating clean EEG to “strong outlier” indicating high likelihood of artifact contamination) \cite{hajra2016brain_vital_signs}.
Channel-wise scores are then aggregated into an epoch-level Signal Quality Index (SQI), with larger SQI indicating lower signal quality. 

FAAR then achieves adaptive thresholding by selecting the rejection threshold automatically from the empirical SQI distribution using a knee-detection algorithm \cite{satopaa2011kneedle}, implemented in the Kneeliverse Python package \cite{Antunes2025}. This yields a data-driven threshold that adapts to recording quality across subjects and sessions, avoids manual recalibration, and removes the cross-validation/optimization overhead of methods such as AutoReject. 

\subsection{Datasets} We evaluate FAAR across 13 publicly available MI datasets accessed through MOABB \cite{aristimunha2025moabb}. The datasets span multiple paradigms, session counts, and recording conditions; a summary is provided in Fig.~\ref{datasets}. When the cross-session analysis is performed in this study, only subjects with at least two recording sessions were retained. We rely on MOABB’s standardized dataset access and protocol definitions to ensure consistent cross-dataset evaluation. 

\begin{figure*}[!htbp]
    \centering
    \includegraphics[width=1\linewidth]{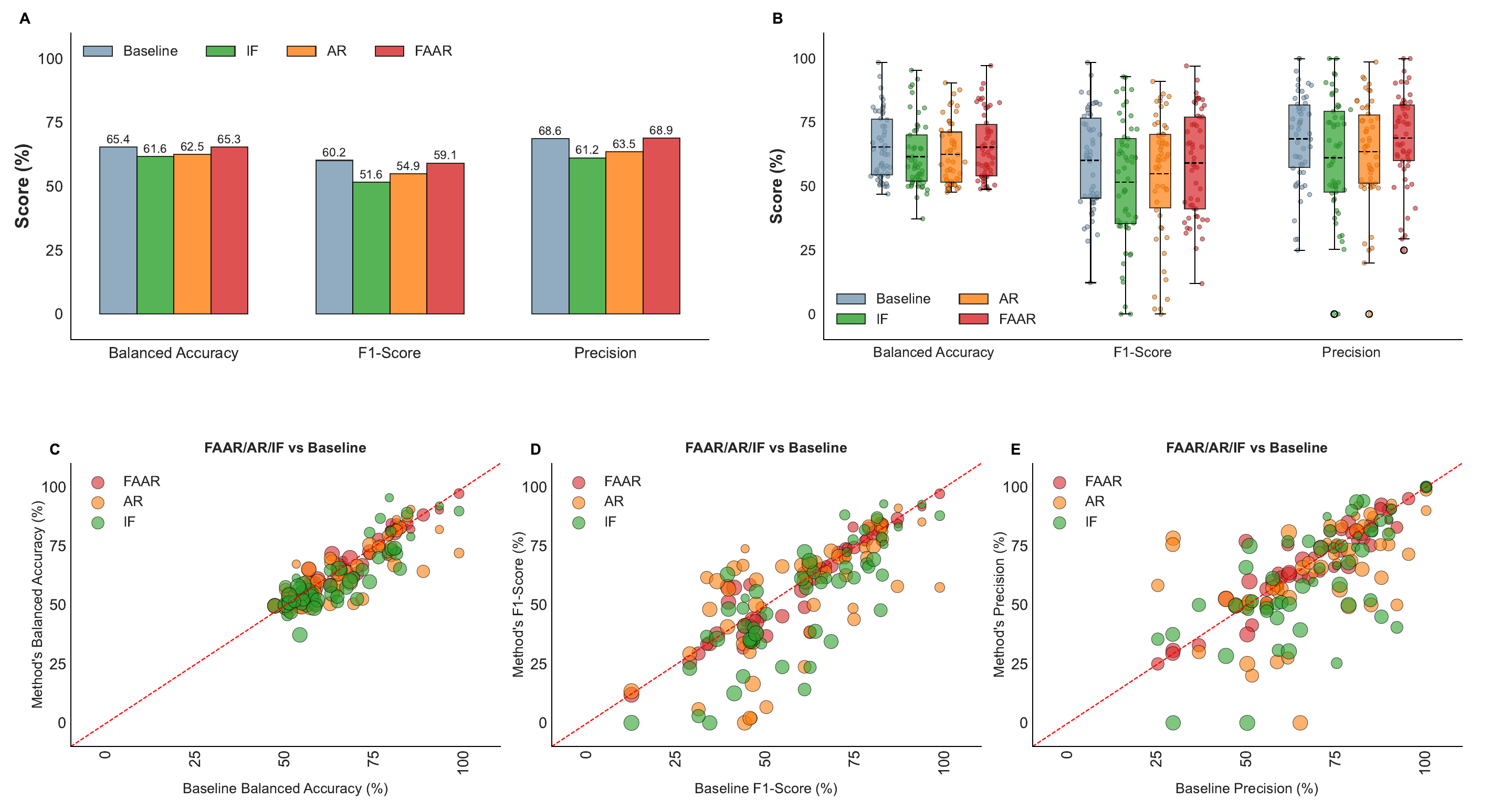}
    \caption{Comparison of artifact rejection methods (AR, IF, and FAAR) against the baseline on the Lee2019\_MI dataset under cross-session evaluation. (A) Mean performance across subjects and (B) subject-level distributions. (C)–(E) Per-subject comparison of FAAR, AR and IF versus baseline for balanced accuracy, F1-score, and precision. In the scatter plots, each point corresponds to a subject, with marker size proportional to $1/\mathrm{ECE}$ (inverse expected calibration error), allowing performance changes to be interpreted jointly with calibration; subjects with larger markers correspond to better-calibrated models. }
    \vspace{-4mm}
    \label{fig:example_detailed_dataset}
\end{figure*}

\subsection{Evaluation Pipeline}

Evaluation of FAAR on downstream MI decoding performance follows a cross-session and cross-subject decoding protocol.
For each dataset, signals are band-pass filtered in the MI band (8--32~Hz) and epoched according to the cognitive event region of interest \cite{aristimunha2025moabb}. 
Artifact handling is applied within each training fold under four compared conditions: no artifact handling (baseline), AutoReject (AR), Isolation Forest (IF), and our proposed FAAR. Rejected epochs are excluded from training/testing.

Decoding is performed with $K$-fold cross-validation, holding out one session (cross-session) or one subject (cross-subject) per fold, training on the remaining data, and averaging scores across folds. To isolate the effect of rejection, all conditions use the same downstream decoder: covariance estimation, tangent-space projection, and logistic regression \cite{barachant2012multiclass_riemannian,chevallier2024MOABB}, a robuststate-of-the-art method according to a recent benchmark. 

Performance is reported with balanced accuracy (BA), F1-score, and precision. To characterize deployment-relevant behavior beyond mean accuracy, we also report: (i) win rate (fraction of subjects improved vs.\ baseline and other methods), (ii) inter-subject variability (standard deviation of balanced accuracy across subjects, averaged per dataset/paradigm), (iii) epoch rejection rate, and (iv) calibration via expected calibration error (ECE), jointly capturing performance, reliability, and the aggressiveness of each rejection strategy. To fairly benchmark the operational cost, we measured single-thread CPU wall-time per second of EEG on a representative dataset (same machine for all methods).  

\begin{figure*}
    \centering
    \includegraphics[width=\linewidth]{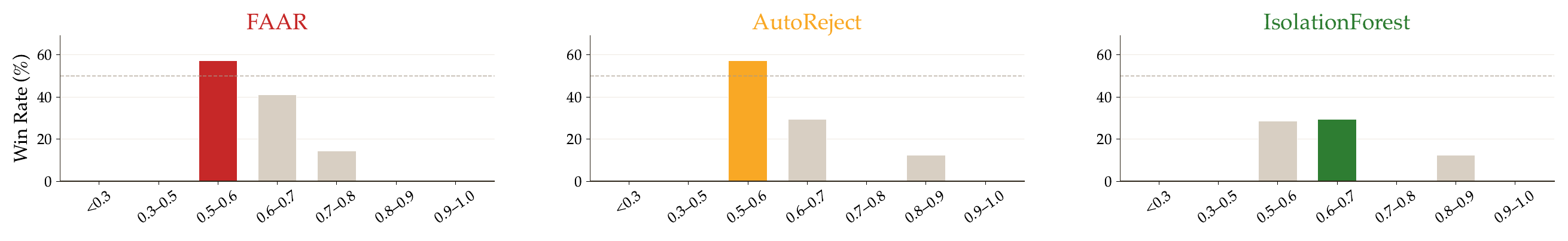}
    \caption{Win rate for Left–Right Imagery: percentage of subjects for whom cleaning improves balanced accuracy (BA), as a function of baseline performance. For baseline BA $<$ 0.6, FAAR and AR improve performance for 57\% of subjects, whereas IsolationForest reaches only 29\%; for baseline BA $\geq$ 0.6, all methods show win rates below 20\%, indicating that artifact cleaning is less likely to benefit already higher-performing subjects. The dashed horizontal line marks the 50\% win-rate reference.}
    \label{fig:winrate_vs_baseline}
    \vspace{-5mm}
\end{figure*}

\begin{figure}[htbp]
    \centering
    \includegraphics[width=1\linewidth]{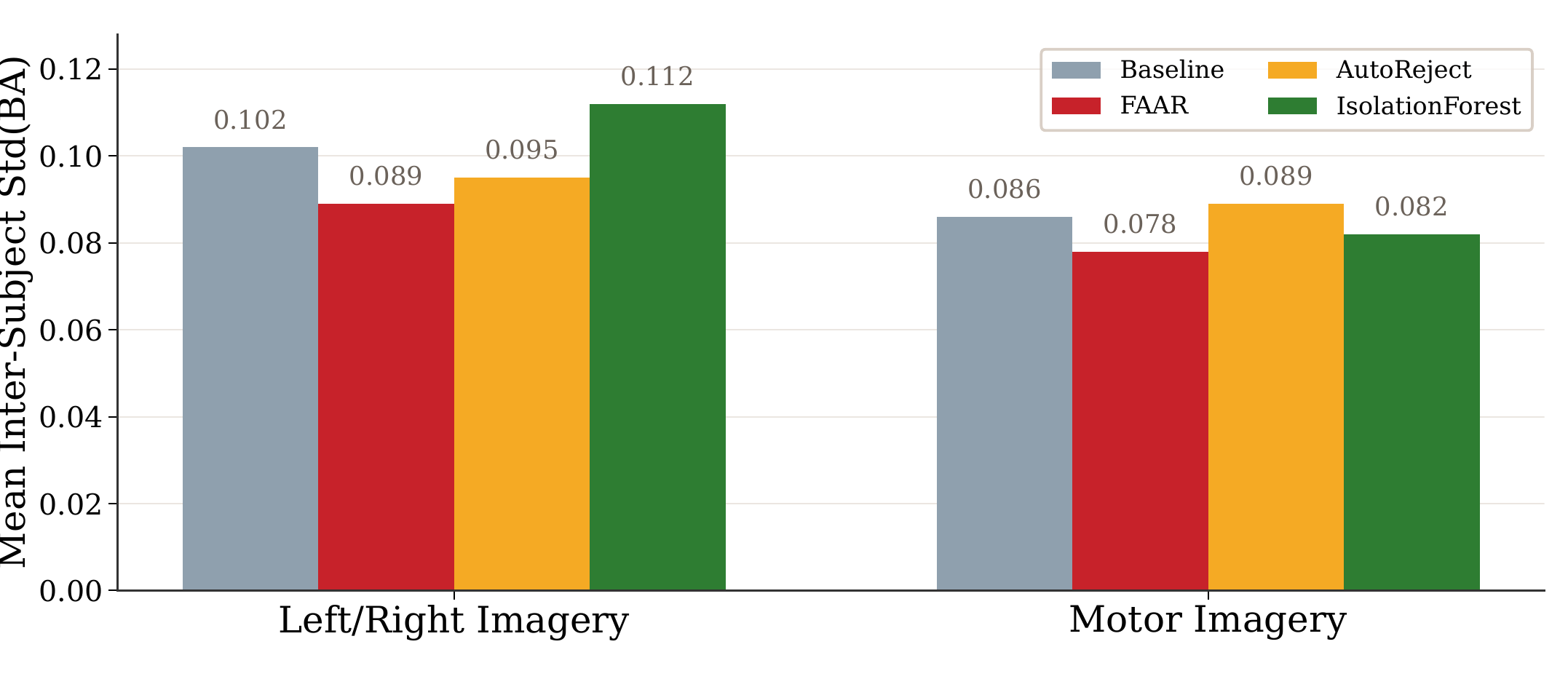}
    \caption{Effect of cleaning on inter-subject variability. Mean inter-subject standard deviation of BA across datasets, reported by paradigm (LR imagery, motor imagery) and method (baseline, FAAR, AR, and IF). Lower values indicate more consistent performance across subjects.}
    \label{fig:inter_subject_variability}
    \vspace{-5mm}
\end{figure}

\begin{figure}[htbp]
    \centering
    \includegraphics[width=1\linewidth]{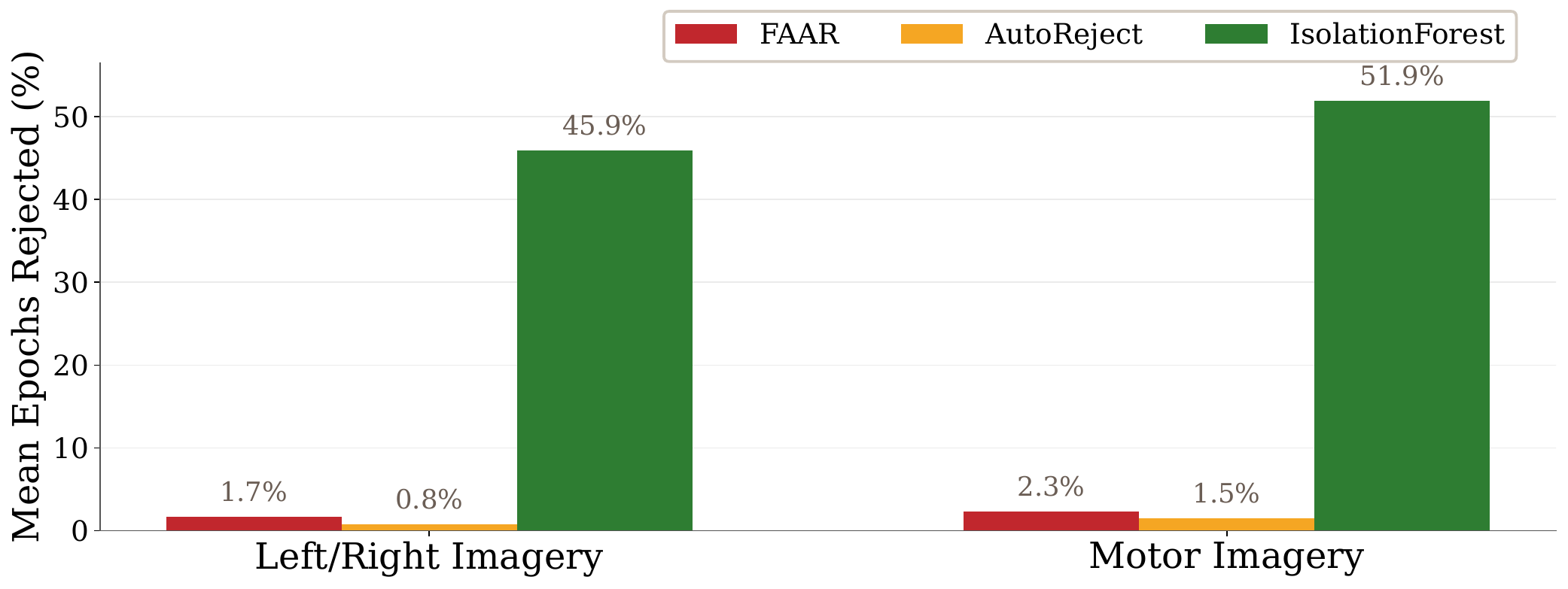}
    \caption{Mean percentage of epochs rejected for LR imagery, motor imagery. IF rejects a substantially larger fraction of data (45-52\%) than FAAR ($\leq$2.3\%) and AutoReject ($\leq$1.5\%).}
    \label{fig:rejection_rates}
\end{figure}

\section{Results}

We report cross-session and cross-subject decoding results across 13 MI datasets (MOABB), comparing four artifact-handling conditions: no rejection (baseline), AutoReject (AR), Isolation Forest (IF), and FAAR.

\noindent \textbf{Aggregate decoding performance.}

A representative example (Lee2019\_MI, cross-session) is shown in Fig.~\ref{fig:example_detailed_dataset}: beyond mean performance, it shows that subject-level dispersion differs markedly across artifact handling methods.

\noindent \textbf{Subject-level regime dependence (win-rate vs.\ baseline).}
Fig.~\ref{fig:winrate_vs_baseline} summarizes the win rate (fraction of subjects improved vs.\ baseline) as a function of baseline BA aggregated across datasets. It shows that artifact rejection does not yield uniform gains in MI-BCI decoding: For low baseline performance (BA $<$0.6), rejection improved balanced accuracy for a majority of subject-level evaluations with FAAR and AR, both reaching approximately 57\% win rate, while IF reached only 2\%. On the contrary, for high baseline performance (BA $\geq$0.6), cleaning more often degrades performance.

\noindent \textbf{Robustness across users.}
Fig.~\ref{fig:inter_subject_variability} reports how artifact handling methods impact inter-subject variability (standard deviation of BA across subjects, averaged per dataset/paradigm). FAAR reduces inter-subject dispersion relative to baseline by $\Delta\sigma_{\text{BA}}=\text{0.008}$ (motor imagery) and $\text{0.013}$ (LR imagery), and is lower than AR/IF by $\text{0.008}/\text{0.013}$ on average.

\noindent \textbf{Rejection rate (operating point).}
Fig.~\ref{fig:rejection_rates} shows that FAAR operates in a conservative regime, rejecting $\leq \text{2.3}\%$ of epochs on average (by paradigm), which is comparable to AR ($\leq \text{1.5}\%$), while IF rejects substantially more (\emph{e.g.}, $\text{45--52}\%$).

\section{Discussion and Future Considerations}
Our results support several deployment-relevant takeaways.

\noindent \textbf{Artifact rejection is subject- and regime-dependent, so it should be used adaptively.}
The win-rate curve (Fig.~\ref{fig:winrate_vs_baseline}) shows that rejection helps mainly in low-baseline/low-SNR regimes, especially for FAAR and AR, and can hurt when baseline decoding is already strong (\emph{e.g.}, by removing usable trials and reducing effective sample size). Practically, rejection should be treated as a \emph{conditional} quality intervention layer (triggered or tuned by the operating regime), not a default pre-processing step. This is a key point for cross-session deployment across subjects where artifact statistics drift over time.

\noindent \textbf{FAAR improves robustness across users by reducing inter-subject variability.}
FAAR consistently reduces subject-to-subject dispersion relative to baseline and typically vs.\ AR/IF (Fig.~\ref{fig:inter_subject_variability}), yielding more consistent MI decoding across participants. This is particularly valuable for cross-session BCI operation and downstream systems where predictable user-level behavior matters as much as marginal mean performance gains.

\noindent \textbf{FAAR, like AR, is conservative and avoids IF's data-loss failure mode.}
FAAR rejects only a small fraction of epochs (Fig.~\ref{fig:rejection_rates}), whereas IF can remove a large share of trials, which reduces effective sample size, and can destabilize training/testing and amplify variance, potentially degrading performance (see Fig.~\ref{fig:example_detailed_dataset}) in high-baseline regimes. By preserving statistical power, and reducing the latency required to accumulate usable trials, FAAR is better suited to data-limited and real-time settings.

\noindent \textbf{FAAR removes threshold calibration burden while remaining lightweight for real-time use.}
FAAR avoids user-specified and dataset-specific threshold calibration by (i) normalizing features to a self-derived reference and (ii) selecting the rejection threshold automatically (knee detection), yielding an adaptive threshold that transfers across subjects/sessions at low compute cost. In contrast, AR relies on cross-validation and optimization, which is heavier. In a runtime benchmark, FAAR, like IF, is real-time compatible (\emph{i.e.}, processing faster than signal duration) and substantially cheaper than AR while typically yielding stronger decoding than IF under comparable efficiency constraints (Fig.~\ref{fig:example_detailed_dataset}).

\noindent \textbf{FAAR provides a repeatable quality decision layer from offline curation to real-time deployment.}
FAAR uses a fixed SQI definition and deterministic, reference-normalized knee thresholding. Therefore, ``clean enough'' is defined consistently during dataset construction, model training, and live inference. This reduces train-deploy mismatch and simplifies integration and monitoring, whereas AR's data-dependent threshold selection and IF's non-reference-anchored behavior can yield less stable rejection behavior across offline vs.\ online contexts.

\noindent \textbf{FAAR is interpretable, diagnosable, and easy to adapt.}
FAAR’s SQI is built from simple analytical features, so rejection decisions can be inspected and explained. FAAR is presented here for artifact rejection, but the same SQI-based mechanism can also be used as a decision-level layer for other artifact-handling strategies, including correction \cite{GEDAI}. This makes it easier to understand why epochs are rejected, corrected, or otherwise flagged, and to adjust the feature set or thresholding strategy when transferring across datasets/devices, a practical advantage over black-box anomaly detectors such as IF and over heavier calibration pipelines such as AR.

\textbf{Limitations.}
This study uses public MI datasets and a standardized tangent-space classifier to isolate rejection effects. Future work should stress-test robustness under controlled noise injections, and extend evaluation to real-world recordings with a higher density of artifacts (\emph{e.g.}, dry electrodes, motion-heavy and uncontrolled settings), where FAAR's artifact rejection may yield clearer gains over the no-cleaning baseline. The present implementation of FAAR does not explicitly exploit spatial structure beyond channel-wise aggregation, which remains an important direction for future work. Additional decoders, including deep learning models, should also be assessed, but the present results already establish clear operating-regime guidance and an operationally efficient rejection baseline for cross-session MI pipelines.

\textbf{Acknowledgements.} This project was supported with compute and storage resources by GENCI at IDRIS under grant 2025-A0201016987 on the A100 and H100 partitions of the Jean Zay supercomputer.

\end{document}